\documentclass[12pt]{article}
\usepackage{amsmath}
\usepackage{graphicx,psfrag,epsf}
\usepackage{epsfig}
\usepackage{epstopdf}
\usepackage{enumerate}
\usepackage{natbib}
\usepackage{algpseudocode}
\usepackage{url} 
\usepackage{hyperref, bm}
\usepackage{amsmath,amssymb}
\usepackage{bbm}
\usepackage{xcolor}
\usepackage{caption}
\usepackage{array}
\usepackage{subcaption}
\usepackage{enumitem}
\usepackage{multirow}
\usepackage{amsfonts,amstext,amsmath,amssymb,amsthm}
\usepackage{mathtools}
\usepackage{algorithm}

\newcolumntype{P}[1]{>{\centering\arraybackslash}p{#1}}
\newcommand{\E}{\mathbb{E}}

\newtheorem{definition}{Definition}

\newtheorem{Example}{Example}

\newtheorem{theorem}{Theorem}
\newtheorem{remark}{Remark}

\newcommand\numberthis{\addtocounter{equation}{1}\tag{\theequation}}

\newcommand{\blind}{0}
\begin{document}

\def\spacingset#1{\renewcommand{\baselinestretch}%
{#1}\small\normalsize} \spacingset{1}


\if0\blind
{
  \title{\bf On the Confidence Intervals in Bioequivalence Studies}
  \author{Kexuan Li\thanks{
    Corresponding author. Email Address: kexuan.li.77@gmail.com}, Susie Sinks, Peng Sun\hspace{.2cm}\\
     Global Analytics and Data Sciences, Biogen, Cambridge, Massachusetts, US.\\
    and \\
     Lingli Yang\\
    Department of Mathematical Sciences, Worcester Polytechnic Institute.}
  \maketitle
} \fi

\if1\blind
{
  \bigskip
  \bigskip
  \bigskip
  \begin{center}
    {\LARGE\bf Title}
\end{center}
  \medskip
} \fi

\bigskip
\begin{abstract}

A bioequivalence study is a type of clinical trial designed to compare the biological equivalence of two different formulations of a drug. Such studies are typically conducted in controlled clinical settings with human subjects, who are randomly assigned to receive two formulations. The two formulations are then compared with respect to their pharmacokinetic profiles, which encompass the absorption, distribution, metabolism, and elimination of the drug. Under the guidance from Food and Drug Administration (FDA), for a size-$\alpha$ bioequivalence test, the standard approach is to construct a $100(1-2\alpha)\%$ confidence interval and verify if the confidence interval falls with the critical region. In this work, we clarify that $100(1-2\alpha)\%$ confidence interval approach for bioequivalence  testing yields a size-$\alpha$ test only when the two one-sided tests in TOST are ``equal-tailed''. Furthermore, a $100(1-\alpha)\%$ confidence interval approach is also discussed in the bioequivalence study.
\end{abstract}

\noindent%

\spacingset{1.3}
\newpage

\section{Motivation}

Bioequivalence studies are a type of clinical trial designed to compare the biological equivalence of two different formulations of a drug. The objective of these studies is to establish that the generic drug demonstrates the same rate and extent of absorption as the reference drug, and that its therapeutic effects are comparable. Such studies are typically conducted in controlled clinical settings with human subjects, who are randomly assigned to receive either the generic drug or the reference drug. The two formulations are then compared with respect to their pharmacokinetic profiles, which encompass the absorption, distribution, metabolism, and elimination of the drug. Acceptance criteria for bioequivalence are generally based on statistical methods that assess the similarities between the pharmacokinetic profiles of the two drug products. These comparisons often involve evaluating the area under the concentration-time curve (AUC) and the maximum concentration (Cmax) of the drug in the bloodstream. By ensuring that both AUC or Cmax values fall within pre-defined equivalence ranges, researchers can determine if the two formulations are bioequivalent.


Suppose we want to access the bioequivalence of two drugs or formulations. Under the guidance of Food and Drug Administration (FDA), the bioequivalence is claimed if a $90\%$ two-sided confidence interval of the geometric mean ratio fall within 80-125$\%$. Under the context of statistical hypothesis testing, to demonstrate bioequivalence, the following significance level $\alpha$ hypothesis testing problem is considered:
\begin{equation}\label{eq:hypothese_ratio}
H_0: \frac{\eta_T}{\eta_R} \leq \delta_L \textrm{ or } \frac{\eta_T}{\eta_R} \geq \delta_U \textrm{ versus } H_a: \delta_L < \frac{\eta_T}{\eta_R}< \delta_U,
\end{equation}
where $\eta_T, \eta_R$ are the population geometric mean of the test product and reference product respectively, $\delta_L, \delta_U$ are the lower and upper error bounds determined by the regulatory. Under the theory of intersection-union test, \cite{westlake1981bioequivalence, schuirmann1981hypothesis, schuirmann1987TOST} proposed two one-sided tests (TOST), which now has been the standard approach to test (\ref{eq:hypothese_ratio}). As its name implies, TOST decomposes the interval hypothesis into two size-$\alpha$ one-side hypothesis and reject $H_0$ if and only if both two one-sided hypothese are rejected. In general, when multiple size-$\alpha$ tests are combined together, the overall size of the combined test is no longer $\alpha$. Fortunately, due to the theory of interaction-union test, the size of TOST is still $\alpha$, which will be discussed in Section \ref{sec:intersection_union_test}. It has also been showed that, for a size-$\alpha$ TOST procedure, it is identical to construct a $100(1-2\alpha)\%$ confidence interval for $\frac{\eta_T}{\eta_R}$ and reject $H_0$ if the confidence interval falls entirely between $\delta_L$ and $\delta_U$. Even though the procedure seems simple, there are still some questions need to clarify, for example:
\begin{itemize}
\item Why is the geometric mean preferred over the arithmetic mean?
\item Why are the BE limitss (0.8, 1.25) not symetric about 1?
\item Why is the product of BE limit equal to 1 ($0.8\times1.25=1$), i.e., symmetric about 1 on the ratio scale?
\item In general, the combination of two size-$\alpha$ tests could not get a size-$\alpha$ test, but why TOST works? Why is a size-$\alpha$ TOST identical to a $100(1-2\alpha)\%$ confidence interval, not a $100(1-\alpha)\%$ confidence interval? Is TOST the most powerful?
\end{itemize}
The purpose of this paper is answer these questions theoretically and provide a clear and in-depth understanding of the bioequivalence study. For example, the geometric mean is preferred in bioequivalence studies because pharmacokinetic parameters, such as AUC and Cmax, typically follow a log-normal distribution. Using the geometric mean ensures that the ratio of the means remains unaffected by extreme values or outliers, leading to a more accurate and robust comparison of the test and reference formulations. The bioequivalence limits (0.8, 1.25) are not symmetric around 1 because they account for potential differences in variability and the possibility of type I or type II errors. The limits are based on log-transformed data, and when back-transformed to the original scale, they yield asymmetric confidence intervals. The asymmetry helps to accommodate the potential differences in variability between the two drug formulations. The product of the BE limits is equal to 1 to ensure that the ratio of the test and reference products' geometric means is unbiased on the ratio scale. When the two limits are multiplied, they effectively cancel out any deviation from the true ratio of 1, allowing for a more accurate comparison of the drug formulations. TOST works because of the intersection-union test theory. While it may not be the most powerful test in all situations, it maintains the overall size-$\alpha$ test by decomposing the interval hypothesis into two size-$\alpha$ one-sided hypotheses and rejecting the null hypothesis only if both one-sided hypotheses are rejected. The intersection-union test theory ensures that the overall type I error rate remains controlled at the desired level ($\alpha$) when combining the two one-sided tests. In general, combining two size-$\alpha$ tests does not guarantee a size-$\alpha$ test because the overall type I error rate may be inflated due to multiple testing. However, the TOST procedure, based on the intersection-union test theory, accounts for this issue and successfully maintains the overall type I error rate at the desired level, making it a suitable method for bioequivalence testing.

The rest of the paper is organized as follows. Section \ref{sec:theoretical_background} provides the theoretical background of bioequivalence study for an univariate PK parameter. We summarize the paper in Section \ref{sec:conclusion} and technical proofs are provided in Appendix.
\section{Theoretical Background} \label{sec:theoretical_background}

In this section, we delve into the theoretical background of bioequivalence studies for a univariate pharmacokinetic (PK) parameter. In a typical pharmacokinetic bioequivalence study, the univariate response variables such as $\log(\textrm{AUC}), \log(\textrm{C}_{\textrm{max}})$ are often assumed to follow a normal distribution, or equivalently, $\textrm{AUC}, \textrm{C}_{\textrm{max}}$ follow a lognormal distribution (\cite{shen2006efficient}, \cite{shen2017checking}). Under FDA guidelines, bioequivalence is treated as a statistical hypothesis testing problem as defined in (\ref{eq:hypothese_ratio}), which states that the population geometric mean ratio for the test and reference products should fall between 80-125$\%$. At first glance, it may seem odd that the geometric mean ratio is used instead of the arithmetic mean difference. Moreover, it might appear strange that 80$\%$ and 125$\%$ are not symmetric about 100$\%$, but $80\%\times125\%=100\%$,  meaning that the lower and upper bounds are symmetric about 100$\%$ on a ratio scale. To explain the reason, we need to take the logarithm of (\ref{eq:hypothese_ratio}) and obtain the following hypothesis testing for the difference:
\begin{equation}\label{eq:hypothese_difference}
H_0: \mu_T - \mu_R \leq \theta_L \textrm{ or } \mu_T - \mu_R \geq \theta_U \textrm{ versus } H_a: \theta_L < \mu_T -\mu_R< \theta_U,
\end{equation}
where $\mu_T=\log(\eta_T), \mu_R=\log(\eta_R), \theta_L=\log(\delta_L), \theta_R=\log(\delta_R)$. The following theorem explain some of the reasons why FDA choose geometric mean ratio as the test statistic and 80-125$\%$ as the BE limits.
\begin{theorem}\label{thm:geometric_mean}
Suppose $X_1, \ldots, X_n$ are independent and identically distributed from normal distribution $X\sim\mathcal{N}(\mu, \sigma^2)$ and $X^*_i = \exp(\mu+\sigma X_i)$, that is, $X^*_1, \ldots, X^*_n$ are independent and identically distributed from lognormal distribution $X^*$ with corresponding mean $\mu$, variance $\sigma^2$ on the log transformed scale. We let $\textrm{GM}(X^*) = (\prod_{i=1}^nX^*_i)^{\frac{1}{n}}$ be the geometric mean of $X^*_1, \ldots, X^*_n$. Then the following statements hold true
\begin{enumerate}[label={(\arabic*)}]
\item $\exp(\bar{X}) = \textrm{GM}(X^*)$;
\item $\textrm{Median}(X^*) = \exp(\mu)$;
\item $\E[\textrm{GM}(X^*)] = \exp(\mu + \frac{\sigma^2}{2n})$.
\end{enumerate}
\end{theorem}
The proof is given in the supplementary material. According to Theorem \ref{thm:geometric_mean}, it is straightforward to obtain the following facts:
\begin{remark}\hfill
\begin{enumerate}[label={(\alph*)}]
\item Since the PK parameters are lognormal distributed, which is right-tailed, it is more natural to compare the median between treatment and reference products rather than the arithmetic mean;
\item According to (1) in Theorem \ref{thm:geometric_mean}, we know that the log-transformed geometric mean of PK data is equivalent to the arithmetic mean of the log-transformed PK data. In other words, comparing the geometric mean ratio between two products is equivalent to comparing the arithmetic mean difference between the log-transformed PK data;
\item According to (2) and (3) in Theorem \ref{thm:geometric_mean}, we can make a conclusion that the geometric mean of PK data is a nearly unbiased estimator of PK median with an error rate of $O(1/n)$. To be more specific, as $n$ approaches infinity, $\E[\textrm{GM}(X^*)] = \textrm{Median}(X^*)$. This implies that, for large sample sizes, the geometric mean of PK data provides an accurate estimate of the PK median.
\end{enumerate}
\end{remark}

Combining everything above, we can identify the advantages of using the geometric mean ratio as the test statistic: (1) The comparison of the geometric mean ratio can be converted to the comparison of the arithmetic mean difference; (2) The geometric mean ratio is a ``good" estimator of the median ratio, which can be naturally applied to the distribution of PK parameters. Follows from the above analysis, we can further clarify the procedure of bioequivalence testing for univariate PK parameter in practice:
\begin{itemize}
\item Step 1: Log-transform the PK data (e.g., AUC and Cmax) to convert the lognormal distribution into a normal distribution.
\item Step 2: Test $H_0: \mu_T - \mu_R \leq \theta_L \textrm{ or } \mu_T - \mu_R \geq \theta_U \textrm{ versus } H_a: \theta_L < \mu_T -\mu_R< \theta_U$, where  $\mu_T, \mu_R$ are the log-transformed PK data from Step 1 and $\theta_L=\log(0.8)=-0.223, \theta_U=\log(1.25)=0.223.$
\item Find a $100(1-2\alpha)\%$ confidence interval for $\mu_T - \mu_R$ and reject $H_0$ if and only if the $100(1-2\alpha)\%$ confidence interval falls in $[\theta_L, \theta_R]$ and the significance level of the corresponding procedure is $\alpha$.
\end{itemize}
It is important to note that after taking the logarithm, the original BE limits (0.8, 1.25) become symmetric about zero, and this fact plays a crucial role in the $100(1-2\alpha)\%$ confidence interval approach. Without the ``equal-tail'' property, the $100(1-2\alpha)\%$ confidence interval approach is no longer valid. This is because the symmetric nature of the limits on the log scale simplifies the comparison of the test and reference products, allowing for a balanced evaluation of bioequivalence that accounts for both type I and type II errors. See Section \ref{sec:two_alpha_CI} for more details.
\subsection{Union-Interaction Tests and Interaction-Union Tests} \label{sec:intersection_union_test}
In the section, we proceed to review the TOST, or two one-sided tests. The TOST involves performing two one-sided tests, one to test if the difference between the treatments or populations is greater than a predetermined non-inferiority margin and another to test if the difference is less than a predetermined equivalence margin. If the results of both tests are significant, it suggests that the treatments or populations are equivalent within the defined margins. Before proceeding, we first review the union-interaction test and intersection-union test, which provide the theoretical foundation for the TOST. Suppose the parameter space of a hypothesis testing is $\Theta = \Theta_0 \bigcup \Theta_0^c$, where $\Theta_0, \Theta_0^c$ are the parameter space of null and alternative hypothesis, respectively. Furthermore, let $R$ be the rejection region of the test, then the power function of the test, as a function of $\theta \in \Theta$, is defined as $\beta(\theta) = \mathbb{P}_{\theta\in\Theta}(T\in R)$, where $T$ is the test statistics. Before moving on, we first clarify two definitions that are commonly confused with one another.: the \emph{size} and the \emph{level} of a test.
\begin{definition}
For $0 \leq \alpha \leq 1$, a test with power function $\beta(\theta)$ is a size-$\alpha$ test if $\sup_{\Theta_0}\beta(\theta) = \alpha$.
\end{definition}
\begin{definition}
For $0 \leq \alpha \leq 1$, a test with power function $\beta(\theta)$ is a level-$\alpha$ test if $\sup_{\Theta_0}\beta(\theta) \leq \alpha$.
\end{definition}
These definitions help distinguish between the size and the level of a test, which are related but distinct concepts. The size of a test is the maximum probability of making a Type I error (i.e., rejecting the null hypothesis when it is true) under the null hypothesis parameter space $\Theta_0$. In contrast, the level of a test refers to the upper bound on the probability of making a Type I error. When the size of a test equals $\alpha$, it is considered a size-$\alpha$ test, whereas when the size of a test is less than or equal to $\alpha$, it is considered a level-$\alpha$ test. For a level-$\alpha$ test, the size of the test may be much less then $\alpha$, and in such case, if we still use $\alpha$ as the significance level, then the test will be less powerful.

Obviously, the null hypothesis in (\ref{eq:hypothese_difference}) can be viewed as the union of two simple null hypothesis $H_{01}: \mu_T - \mu_R \leq \theta_L$ and $H_{02}: \mu_T - \mu_R \geq \theta_U$. Not only in bioequivalence study, but also in some other applications, such complicated hypotheses can be developed from tests for simpler hypotheses. These hypothesis can be generalized from the so called \emph{intersection-union tests} (IUT) and \emph{union-intersection test} (UIT), which play an important role in bioequivalence testing and multiple comparison procedures.
\begin{definition}
Assume a hypothesis $H_0: \theta\in \Theta_0 \textrm{ versus } H_a:\theta\in \Theta^c_0$ with reject $R$. A family of hypotheses $H_{i0}: \theta\in\Theta_{i}$ versus $H_{ia}: \theta\in\Theta^c_{i}$ for $i=1,\ldots,k$ with corresponding rejection region $R_i$ is said to obey the intersection-union principle if
\begin{equation}\label{eq:intersection_union}
H_0: \theta\in\bigcup_{i=1}^k\Theta_i, \textrm{ and } H_a: \theta\in\bigcap_{i=1}^k\Theta^c_i,
\end{equation}
and is said to obey the union-intersection principle if
\begin{equation}\label{eq:union_intersection}
H_0: \theta\in\bigcap_{i=1}^k\Theta_i, \textrm{ and } H_a: \theta\in\bigcup_{i=1}^k\Theta^c_i.
\end{equation}
\end{definition}
It is straightforward to show that the rejection region of IUT is $\bigcap_{i=1}^kR_i$. The logic behind this is simple. $H_0$ is false if and only all of the $H_{0i}, i=1, \ldots, k$ are false. So rejecting $H_0: \theta\in\bigcup_{i=1}^k\Theta_i$ is equivalent as rejecting each individual $H_{0i}: \theta \in \Theta_i$. Similarly, in terms of UIT, using the same logic, if any one of the hypotheses $H_{0i}$ is rejected, then $H_0$ must also be rejected, and only if each of $H_{0i}$ is accepted will the intersection $H_0$ be accepted. Thus, the rejection region of UIT is $\bigcup_{i=1}^kR_i$. Clearly, the hypothesis testing in \ref{eq:hypothese_difference} belongs to IUT. Now, a natural question rises: given the significance of each individual $H_{0i}$, what is the significance level of IUT? The following theorem gives an upper bound for the size of the IUT.
\begin{theorem}\label{thm:IUT_level}
Let $\alpha_i$ be the size of the test of $H_{0i}$ with rejection region $R_i$. Then the IUT with rejection region $R=\bigcap_{i=1}^kR_i$ is a level $\alpha=\sup_{i=1, \ldots,k}\alpha_i$ test.
\end{theorem}
It should be remarked that Theorem \ref{thm:IUT_level} only shows that the level of an IUT is $\alpha=\sup_{i=1, \ldots,k}\alpha_i$, that is, an upper bound for the size of the IUT. In fact, the size of the IUT could be much less than $\alpha$. The following theorem provides the conditions under which the size of the IUT is exactly $\alpha$, ensuring that the IUT is not too conservative.
\begin{theorem}\label{thm:IUT_size}
Consider testing $H_0: \theta\in \bigcup_{j=1}^k\Theta_j$ and let $R_j$ be the rejection region such that the level of $H_{0j}$ is $\alpha$. Suppose that for some $i=1, \ldots, k$, there exists a sequence of parameters, $\theta_l\in \Theta_i, l = 1, 2, \ldots, $ such that
\begin{enumerate}[label={(\arabic*)}]
\item $\lim_{l\rightarrow \infty}\mathbb{P}_{\theta_l}(\bm{X}\in R_i) = \alpha$,
\item for each $j=1, \ldots,k, j\neq i, \lim_{l\rightarrow \infty}\mathbb{P}(\bm{X}\in R_i) = 1$.
\end{enumerate}
Then, the IUT with rejection region $R=\bigcap_{j=1}^kR_j$ is a size $\alpha$ test.
\end{theorem}
\subsection{Two One-sided Tests (TOST)}
We are now in a position to formally introduce the method for bioequivalence testing in (\ref{eq:hypothese_difference}). Albeit good theoretical guarantees for IUT, \cite{westlake1981bioequivalence, schuirmann1981hypothesis, schuirmann1987TOST} proposed the following so-called two one-sided tests (TOST), which has now been one of the standard procedures of (\ref{eq:hypothese_difference}). As its name implies, TOST consists two one-sided hypotheses
\begin{align}\label{eq:TOST_1}
H_{01}: \mu_T - \mu_R \leq \theta_L \textrm{ versus } H_{a1}: \mu_T -\mu_R > \theta_L \end{align}
and
\begin{align}\label{eq:TOST_2}
H_{02}: \mu_T - \mu_R \geq \theta_U \textrm{ versus } H_{a2}: \mu_T -\mu_R < \theta_U.
\end{align}
The $H_0: \mu_T - \mu_R \leq \theta_L \textrm{ or } \mu_T - \mu_R \geq \theta_U$ in (\ref{eq:hypothese_difference}) can be expressed as the union of $H_{01}$ and $H_{02}$. This procedure establishes bioequivalence at significance level $\alpha$ if both $H_{01}$ and $H_{02}$ are rejected at level-$\alpha$. The rationale underlying is simple. If one may conclude that $\mu_T - \mu_R > \theta_L$ and also
$\mu_T - \mu_R < \theta_U$, then it has in effect been concluded that $\theta_L< \mu_T - \mu_R < \theta_U$. In practice, size-$\alpha$ Student's $t$-test is used for (\ref{eq:TOST_1}) and (\ref{eq:TOST_2}) and $H_0$ is rejected if
\begin{equation}
T_L = \frac{(\bar{X}_T - \bar{X}_R) - \theta_L}{\hat{\sigma}_{\Delta\bar{X}}} > t_{1-\alpha, r}, \quad T_U = \frac{(\bar{X}_T - \bar{X}_R) - \theta_U}{\hat{\sigma}_{\Delta\bar{X}}} < -t_{1-\alpha, r},
\end{equation}
where $\bar{X}_T, \bar{X}_R$ are the average of log-transformed PK data for treatment/reference product, and $\hat{\sigma}_{\Delta\bar{X}}=S_p\sqrt{(\frac{1}{n_T} + \frac{1}{n_R})}$ is the standard error of $\bar{X}_T-\bar{X}_R$ and $S_p^2=\frac{1}{n_T+n_R-2}[(n_T-1)S_T^2 + (n_R-1)S_R^2]$, with $S_T, S_R$ being the standard deviation of two groups, and $t_{1-\alpha, r} = \mathbb{P}(X \leq t_{1-\alpha, r}) = 1-\alpha$ is the critical value from $t$ distribution with degree of freedom $r=n_T + n_R -2$.  In addition, it should be mentioned that the size for each individual $t$-test $H_{01}, H_{02}$ is $\alpha$, not $\alpha/2$. There is no need for multiplicity adjustment in
testing each the two one-sided null hypotheses for a univariate PK parameter. Suppose the size for each $t$-test is $\alpha$, since TOST belongs to IUT, from Theorem \ref{thm:IUT_level}, we know the level of TOST is $\alpha$, that is, the size is at most $\alpha$. To prove the size of TOST is exactly $\alpha$, we need to check the conditions of Theorem \ref{thm:IUT_size}. First, consider a parameter point ${\theta_l}=\mu_T - \mu_R = \theta_U$, then we have
\begin{equation*}
\lim_{l\rightarrow \infty}\mathbb{P}_{\theta_l}(\bm{X}\in R_2) = \mathbb{P}_{\theta_U}(\bm{X}\in R_2) = \mathbb{P}_{\theta_U}(T_U < -t_{1-\alpha, r})=\alpha,
\end{equation*}
thus the first condition in Theorem \ref{thm:IUT_size} holds true. Furthermore,
\begin{equation*}
\lim_{l\rightarrow \infty}\mathbb{P}_{\theta_l}(\bm{X}\in R_1) = \mathbb{P}_{\theta_U}(\bm{X}\in R_1)= \mathbb{P}_{\theta_U}(T_L > t_{1-\alpha, r}) \rightarrow 1, \textrm{ as }\sigma^2 \rightarrow 0,
\end{equation*}
therefore, the second condition in Theorem \ref{thm:IUT_size} also remains valid, from where, we summarize the above analysis to the following theorem
\begin{theorem}\label{thm:TOST_size}
Suppose the size of each individual test in TOST is $\alpha$, then the size of TOST equals $\alpha$ exactly.
\end{theorem}

To compute the power of TOST, \cite{phillips1990power} derived the explicit form of the power, as a function of $(\mu_T, \mu_R, n_T, n_R, \sigma^2, \theta_L, \theta_U)$ given by
$$
\begin{aligned}
\textrm{Power}\left(\mu_T, \mu_R, n_T, n_R, \sigma^2, \theta_L, \theta_U\right) &=Q_{r}\left(-t_{1-\alpha, r}, \frac{\mu_T-\mu_R-\theta_{U}}{\sigma\sqrt{\frac{1}{n_T}+\frac{1}{n_T}}} ; 0 \frac{\left(\theta_{U}-\theta_{L}\right) \sqrt{r}}{2 \sigma\sqrt{\frac{1}{n_T}+\frac{1}{n_T}} t_{1-\alpha, r}}\right) \\
&\quad-Q_{r}\left(t_{1-\alpha, r}, \frac{\mu_T-\mu_R-\theta_{L}}{\sigma\sqrt{\frac{1}{n_T}+\frac{1}{n_T}}} ; 0 \frac{\left(\theta_{U}-\theta_{L}\right) \sqrt{r}}{2 \sigma\sqrt{\frac{1}{n_T}+\frac{1}{n_T}} t_{1-\alpha, r}}\right)
\end{aligned}
$$
where $$
Q_{v}(t, \delta ; a, b)=\frac{\sqrt{2 \pi}}{\Gamma\left(\frac{v}{2}\right) \cdot 2^{\frac{v-2}{2}}} \int_{a}^{b} \Phi\left(\frac{t x}{\sqrt{v}}-\delta\right) x^{v-1} \phi(x) d x, r = n_1 + n_2 - 2
$$
is referred as the Owen's $Q$ function and, $\Phi(\cdot), \phi(\cdot)$ are the cdf and pdf of standard normal distribution, respectively.

In the previous sections, the theoretical foundations of the TOST method for bioequivalence testing have been established. However, the relationship between the TOST and $100(1 - \alpha)\%$ and $100(1 - 2\alpha)\%$ confidence intervals is not yet clear. In the next section, we will explore the connections between the TOST and these two types of confidence intervals, shedding light on their similarities, differences, and implications for bioequivalence testing. Understanding the connections between the TOST and different confidence intervals will provide insights into how these statistical methods relate to each other, and how they can be effectively applied in bioequivalence testing for univariate pharmacokinetic parameters. This will enhance the interpretation and application of these methods in practice, contributing to more accurate and reliable conclusions about bioequivalence.
\subsection{Confidence Sets}
Intuitively, a hypothesis is associated with an equivalent confidence interval approach. Traditionally, bioequivalence is claimed if the $100(1-2\alpha)\%$ two-sided confidence interval of the geometric mean ratio, or equivalently, the arithmetic mean difference of the log-transformed data,  falls within the BE limits. The use of a $100(1 - 2\alpha)\%$ confidence interval for bioequivalence testing, rather than a $100(1 - \alpha)\%$ confidence interval, has been a subject of debate among statisticians. The relationship between the size-$\alpha$ TOST procedure and the $100(1 - 2\alpha)\%$ confidence interval approach has been questioned by some researchers, who argue that the similarity is based more on an algebraic coincidence than a true statistical equivalence. \cite{brown1995optimal} suggested that the association between the TOST procedure and the $100(1 - 2\alpha)\%$ confidence interval approach may be somewhat of a fiction, while \cite{berger1996bioequivalence} pointed out that using a $100(1 - 2\alpha)\%$ confidence interval for bioequivalence testing can be conservative and may only work in specific cases. Other discussions on the use of the $100(1 - 2\alpha)\%$ confidence interval procedure in bioequivalence testing can be found in works by \cite{westlake1976symmetrical},\cite{westlake1981bioequivalence}, \cite{rani2004bioequivalence}, \cite{choi2008survey}. These discussions highlight the ongoing debate about the appropriateness of using a $100(1 - 2\alpha)\%$ confidence interval for bioequivalence testing, and the importance of understanding the underlying statistical concepts and assumptions in order to make informed decisions about the most suitable methods for a given study.
\subsubsection{\boldmath{$100(1-\alpha)\%$} Confidence Interval}
As we mentioned before, there are many different formulations of the bioequivalence hypothesis that lead to alternate tests and confidence intervals. In this section, we will discuss a $100(1-\alpha)\%$ confidence interval approach that corresponds exactly a size-$\alpha$ TOST. It is well-known that there is a closed relationship between level-$\alpha$ hypothesis and $100(1-\alpha)\%$ confidence set, that is, rejecting $H_0$ if and only if the intersection of the $100(1-\alpha)\%$ confidence set and the null hypothesis is empty. We summarize the property into the following theorem
\begin{theorem} \label{thm:confidence_set_test}
Let $\Theta$ be the parameter space. For each $\theta_0\in\Theta$, let $T_{\theta_0}$ be a test statistic for $H_0: \theta=\theta_0$ with significance level $\alpha$ and acceptance region $A(\theta_0)$, then the set $C(\bm{X})=\{\theta: \bm{X}\in A(\theta)\}$ is a level-$\alpha$ confidence set for $\theta$.
\end{theorem}

In other words, suppose $[L(\bm{X}), R(\bm{X})]$ is a confidence interval of parameter $\theta \in \Theta$ with condidence coefficient equal to $1-\alpha$, that is, $\inf_\theta\mathbb{P}_{\theta\in\Theta}(\theta\in[L(\bm{X}), R(\bm{X})])=1-\alpha$. Consider the hypothesis testing $H_0: \theta\in\Theta_0$ verse $H_a: \theta\in\Theta_0^c$. Then from Theorem \ref{thm:confidence_set_test}, we know that the test that rejects $H_0$ if and only if $[L(\bm{X}), R(\bm{X})]\bigcap \Theta_0^c = \O$ is a level-$\alpha$ test. In this section, we will apply Theorem \ref{thm:confidence_set_test} to show that a size-$\alpha$ TOST is associated with a $100(1-\alpha)\%$ confidence interval.
\begin{theorem} \label{thm:confidence_interval_alpha}
Consider the hypothesis testing $H_0: \mu_T - \mu_R \leq \theta_L \textrm{ or } \mu_T - \mu_R \geq \theta_U \textrm{ versus } H_a: \theta_L < \mu_T -\mu_R< \theta_U$ in (\ref{eq:hypothese_difference}) and define the following confidence interval for $\mu_T-\mu_R$
\begin{equation} \label{eq:CI_alpha}
[L(\bm{X}), U(\bm{X})]=\left[\min\{0, (\bar{X}_T - \bar{X}_R) - t_{1-\alpha, r}\hat{\sigma}_{\Delta\bar{X}}\}, \max\{0, (\bar{X}_T - \bar{X}_R) + t_{1-\alpha, r}\hat{\sigma}_{\Delta\bar{X}}\}\right],
\end{equation}
where
\begin{equation}
\hat{\sigma}_{\Delta\bar{X}}=S_p\sqrt{\frac{1}{m}+\frac{1}{n}}, S_p=\sqrt{\frac{(m-1)S_{X_T}^2+(n-1)S_{X_R}^2}{m+n-2}}, r=m+n-2,
\end{equation}
and $S_{X_T}, S_{X_R}$ are the sample standard derivation of the two groups. Then the following two statements hold true
\begin{enumerate}[label={(\arabic*)}]
\item If $\mu_T - \mu_R \neq 0$, then the coverage probability of $[L(\bm{X}), U(\bm{X})]$ is $100(1-\alpha)\%$, otherwise, the coverage probability equals 1.
\item $[L(\bm{X}), U(\bm{X})]$ is associated with the size-$\alpha$ TOST for (\ref{eq:hypothese_difference}).
\end{enumerate}
\end{theorem}
Following Theorem \ref{thm:confidence_interval_alpha}, we could make following remarks.
\begin{remark}\hfill
\begin{enumerate}[label={(\alph*)}]
\item From (1) in Theorem \ref{thm:confidence_interval_alpha}, we know $[L(\bm{X}), U(\bm{X})]$ is a level-$\alpha$ confidence interval.
\item If $[L(\bm{X}), U(\bm{X})]$ contains zero, then it is identical to the $100(1-2\alpha)\%$ confidence interval which will be introduced in the next section.
\item From Bayesian point of view, if the prior distribution of $\mu_T-\mu_R$ is noninformative, that is, $\pi(\mu_T-\mu_R) \propto 1$, then the posterior credible probability of $[L(\bm{X}), U(\bm{X})]$ is exactly $(1-2\alpha)$ for $-t_{1-\alpha, r}\leq \mu_T-\mu_R\leq t_{1-\alpha, r}$ and converges to $(1-\alpha)$ as $|\mu_T-\mu_R|\rightarrow \infty$.
\end{enumerate}
\end{remark}

\subsubsection{\boldmath{$100(1-2\alpha)\%$} Confidence Interval} \label{sec:two_alpha_CI}
Follow the guidance of FDA, in practice, bioequivalence is claimed if the $100(1-2\alpha)\%$ two-sided confidence interval of the geometric mean ratio falls within the BE limits. However, unfortunately, from Theorem \ref{thm:confidence_set_test}, we could only see the connection between TOST and $100(1-\alpha)\%$ confidence interval. The reason why $100(1-2\alpha)\%$ confidence interval yields a size-$\alpha$ test is still unclear. In this section, owing to the ``equal-tail'' property, we will see that is not only an ``algebraic coincidence'', but also theoretically guaranteed. In general, the size of a test associated with a $100(1-2\alpha)\%$ confidence interval is not $\alpha$. For example, consider the following $100(1-2\alpha)\%$ confidence interval
\begin{equation}\label{eq:CI_2alpha}
[L(\bm{X}), U(\bm{X})]=\left[(\bar{X}_T - \bar{X}_R) - t_{1-\alpha_1, r}\hat{\sigma}_{\Delta\bar{X}}, (\bar{X}_T - \bar{X}_R) + t_{1-\alpha_2, r}\hat{\sigma}_{\Delta\bar{X}}\right],
\end{equation}
where $\alpha_1 + \alpha_2 = 2\alpha$. It is obvious that $[L(\bm{X}), U(\bm{X})]$ defined above is a $100(1-2\alpha)\%$ confidence interval. As $\alpha_1 \rightarrow 0$, the confidence interval reduces to $\left(-\infty, (\bar{X}_T - \bar{X}_R) + t_{1-2\alpha, r}\hat{\sigma}_{\Delta\bar{X}}\right]$, and the size associated with this confidence interval is $2\alpha$, not $\alpha$. In fact, the lower confidence interval $[L(\bm{x}), \infty)=[(\bar{X}_T - \bar{X}_R) - t_{1-\alpha_1, r}\hat{\sigma}_{\Delta\bar{X}}, \infty)$ in (\ref{eq:CI_2alpha}) defines a size-$\alpha_1$ test, and similarly, the upper confidence interval $(-\infty, U(\bm{x})] = (\infty, (\bar{X}_T - \bar{X}_R) + t_{1-\alpha_2, r}\hat{\sigma}_{\Delta\bar{X}}]$ in (\ref{eq:CI_2alpha}) defines a size-$\alpha_2$ test. From the IUT theory in Theorem \ref{thm:IUT_level} and Theorem \ref{thm:IUT_size}, we know the size of the defined by $[L(\bm{X}), U(\bm{X})]$ in (\ref{eq:CI_2alpha}) is $\max\{\alpha_1, \alpha_2\}$, which equals $\alpha$ only if $\alpha_1 = \alpha_2$. We summarize the above analysis into the following theorem.
\begin{theorem}\label{thm:confidence_interval_2alpha}
Let $\bm{X}$ be a random sample from a probability distribution with statistical parameter $\theta$. Suppose $[L(\bm{X}), \infty)$ is a $100(1-\alpha_1)\%$ lower confidence interval for $\theta$ and $(-\infty, U(\bm{X})]$ is a $100(1-\alpha_2)\%$ upper confidence interval for $\theta$ and  Then $[L(\bm{x}), U(\bm{x})]$ is a $100(1-\alpha_1-\alpha_2)\%$ confidence interval for $\theta$.
\end{theorem}
\begin{remark}
The above theorem reveals the importance of ``equal-tail'', that is $\alpha_1=\alpha_2$, without this property, the TOST will not yields a size-$\alpha$ test. In other words, the total uncertainty of $2\alpha$ should be spent half above and half below the observed mean. Moreover, this theorem also provides another reason why $\theta_L, \theta_U$ in (\ref{eq:hypothese_difference}) should be symmetric about zero. Otherwise, for example, if $\theta_L + \theta_U < 0$, it is reasonable to put more weights on $\alpha_1$ than $\alpha_2$, in which case, the size of $100(1-2\alpha)\%$ procedure is no longer $\alpha$.
\end{remark}
\subsection{More Powerful Tests}
From the previous analysis, we know that there are various procedures for testing (\ref{eq:hypothese_difference}), and a natural question rises, which one is better? Unfortunately, even though FDA suggests using $100(1-2\alpha)\%$ to test bioequivalence, the procedure is not the best. In fact, it should be remarked that TOST has been criticized by many authors. On the one hand, TOST yields a biased test, on the other hand, the power of TOST is quite low. In practice, we would like a hypothesis test to be more likely to reject $H_0$ if $\theta \in \Theta_0^c$ than if $\theta \in \Theta_0$, that is, $\beta(\theta_2) \geq \beta(\theta_1)$ for every $\theta_1\in \Theta_0$ and $\theta_2\in \Theta_0^c$. The following definition summarize the tests summarizing this property.
\begin{definition}
Let $\alpha$ be the significance level. A hypothesis test $H_0: \theta \in \Theta_0 \textrm{ versus } H_a: \theta \in \Theta_0^c$ with power function $\beta(\cdot)$ is said to be unbiased if and only if
\begin{equation*}
\beta(\theta) \leq \alpha, \theta \in \Theta_0 \textrm{ and } \beta(\theta) \geq \alpha \in \Theta_0^c.
\end{equation*}
\end{definition}
Unfortunately, even using a $100(1-2\alpha)\%$ confidence interval approach, TOST is still a biased test. Furthermore, it has also been shown that TOST is conservative and inefficient under asymptotic setup. Much attention has been paid to equivalence testing problem in (\ref{eq:hypothese_difference}) for decades, to mention but a few, \cite{anderson1983new}, \cite{brown1997unbiased}, \cite{martin2004asymptotical}, \cite{choi2008survey}, \cite{fogarty2014equivalence}, \cite{pesarin2016union}. See \cite{wellek2002testing}, \cite{meyners2012equivalence} for a comprehensive review. Even though many methodologies have been proposed to test bioequivalence, the theoretical results are limited. Except some special models, like normal distribution with known variance, no finite sample optimality theory is available for tests of equivalence. To the best of our knowledge, the work by \cite{romano2005optimal} is the first theoretical result where the asymptotic optimality theory is established. Since the optimality theory is beyond the scope of the paper, we only show two examples to give the author some intuitions about optimal testing of equivalence.
\begin{Example}[Normal Distribution with Known Variance] Suppose $X_1, \ldots, X_n$ are i.i.d. $\mathcal{N}(\mu, \sigma^2)$, where $\sigma$ is known. Consider the hypothesis testing: $H_{0}: |\mu| \geq \theta \textrm{ versus } H_{a}: |\mu| < \theta$. Then the uniformly most powerful level-$\alpha$ test is rejecting $H_0$ is $n^{1/2}|\bar{X}_n| \leq \psi(\alpha, n^{1/2}\theta, \sigma)$, where $\psi(\alpha, \theta, \sigma)$ satisfies
\begin{align*}
\phi(\frac{\psi - \theta}{\sigma}) - \phi(\frac{-\psi - \theta}{\sigma}) = \alpha,
\end{align*}
and $\phi(\cdot)$ is the cdf of standard normal distribution.
\end{Example}
However, the uniformly most powerful test does not exist is $\sigma$ is unknown. The next example generalize the results to the exponential family.
\begin{Example}[One-parameter Exponential Family] Suppose $X_1, \ldots, X_n$ are i.i.d. generated from the following one-parameter exponential family
\begin{align*}
f_\theta(x) =\exp\{\eta(\theta)Y(x) - \xi(\theta)\}h(x),
\end{align*}
where $\eta(\theta)$ is an increasing function of $\theta$ only, $h(x)$ is a function of $x$ only, and $Y(\cdot)$ has monotone likelihood ratio in $(X_1, \ldots, X_n)$. Then the uniformly most powerful level-$\alpha$ test for the equivalence hypothesis testing $H_0: \theta\leq\theta_1 \textrm{ or } \theta\geq\theta_2$ versus $H_1: \theta_1\le\theta\le\theta_2$ is
\begin{equation*}
  T(X_1, \ldots, X_n) =
    \begin{cases}
      1 & \text{if } c_1<Y(X_1, \ldots, X_n)<c_2\\
      \gamma_i & \text{if } Y(X_1, \ldots, X_n)=c_i, i=1,2\\
      0 & \text{if } Y(X_1, \ldots, X_n)<c_1 \text{ or } Y(X_1, \ldots, X_n)>c_2,
\end{cases}
\end{equation*}
where $\gamma_1, \gamma_2, c_1, c_2$ are determined by $\beta(\theta_1) = \beta(\theta_2)=\alpha$.
\end{Example}

\section{Conclusion}\label{sec:conclusion}
The paper discusses the significance of bioequivalence studies, which compare the biological equivalence of two different formulations of a drug. These studies are conducted in clinical settings with human subjects to compare the pharmacokinetic profiles of the two formulations. Acceptance criteria for bioequivalence are generally based on statistical methods that evaluate the similarities between the two formulations. The paper explains the theory of intersection-union tests and how they can be used to determine bioequivalence. The TOST (two one-sided tests) approach is the standard approach to bioequivalence testing and has been shown to maintain a size-$\alpha$ test by rejecting the null hypothesis only if both one-sided hypotheses are rejected. The paper provides insights into the theory behind bioequivalence testing, including the use of geometric mean, the non-symmetry of bioequivalence limits around 1, and the product of bioequivalence limits equaling 1. The paper also discusses the connection between the $100(1-2\alpha)\%$ confidence interval approach and $100(1-\alpha)\%$ confidence interval approach.
\clearpage
\bibliographystyle{apalike}
\bibliography{reference}
\newpage
\appendix
\section{Proof of Theorem \ref{thm:geometric_mean}}
\begin{proof}
The proof of (1) can be accomplished by direct calculations. It can be shown that
\begin{align*}
\exp(\bar{X}) &= \exp(n^{-1}\sum_{i=1}^nX_i)=\exp\{\log[\exp(\sum_{i=1}^nX_i/n)]\} = \exp\{\log[\prod_{i=1}^n\exp(X_i/n)]\} \\
\quad &= (\prod_{i=1}^n\exp(X_i))^{\frac{1}{n}}=(\prod_{i=1}^nX^*)^{\frac{1}{n}}=\textrm{GM}(X^*).
\end{align*}
To prove (2), we let $F_{X^*}(x^*)=\frac{1}{2}\left[1+\textrm{erf}\left(\frac{\log(x^*)-\mu}{\sqrt{2}\sigma}\right)\right]$ be the cumulative distribution function of the lognormal distribution, where $\textrm{erf}(x^*)=\frac{2}{\pi} \int_0^{\frac{\pi}{2}} \exp \left(-\frac{x^{*2}}{\sin ^2 \theta}\right) d \theta$ is the complementary error function. For simplicity, we let the median of $X^*$ equals $\varrho$, that is, $\textit{Median}(X^*) = \varrho$. Then, from the definition of median, we have $F_{X^*}(\varrho)=0.5$, thus $\varrho=F_{X^*}^{-1}(0.5)$. Therefore, it follows that
\begin{equation*}
\varrho=F_{X^*}^{-1}(0.5)=\left.\exp \left[\sigma \sqrt{2} \cdot \operatorname{erf}^{-1}(2 p-1)+\mu\right]\right\vert_{p=0.5},
\end{equation*}
where $\operatorname{erf}^{-1}(\cdot)$ is the inverse of the complementary error function satisfying $\operatorname{erf}^{-1}(0)=0$, thus, the median of $X^*$ equals $\exp(\mu)$. In fact, it can be also seen that logarithm function is a monotonous increasing, therefore, the median of $X^*$ is the exponential of the median of $\log(X^*)$. Since $\log(X^*)$ is normal distributed, thus, median of $X^*$ equals $\exp(\mu)$.

Let $f(x^*) = \frac{1}{x^*\sigma\sqrt{2\pi}}\exp\left(-\frac{(\ln(x^*)-\mu)^2}{2\sigma^2}\right)$ be the pdf of lognormal distribution. Consider the expected value of $X_i^{*\frac{1}{n}}$, $\E [X_i^{*\frac{1}{n}}]$. It follows that
\begin{align*}
\E [X_i^{*\frac{1}{n}}] &= \int_0^\infty\frac{1}{x^*\sigma\sqrt{2\pi}}x^{*\frac{1}{n}}\exp\left(-\frac{(\ln(x^*)-\mu)^2}{2\sigma^2}\right)dx \\
&=\int_{-\infty}^{\infty} \frac{1}{\sigma \sqrt{2 \pi}} \frac{1}{\exp (t)} \exp (t)^{1 / n} \exp \left(-\frac{(t-\mu)^2}{2 \sigma^2}\right) \exp (t) d t \numberthis \label{eq:change_of_variable}\\
&=\int_{-\infty}^{\infty} \frac{1}{\sigma \sqrt{2 \pi}} \exp \left(-\frac{\left\{(t-\mu)^2 n-2 \sigma^2 t\right\}}{2 n \sigma^2}\right) d t \\
&=\int_{-\infty}^{\infty} \frac{1}{\sigma \sqrt{2 \pi}} \exp \left(-\frac{\left\{n\left(t-\mu-\frac{\sigma^2}{n}\right)^2-\frac{\sigma^4}{n}-2 \sigma^2 \mu\right\}}{2 n \sigma^2}\right) d t\\
&=\int_{-\infty}^{\infty} \frac{1}{\sigma \sqrt{2 \pi}} \exp \left(-\frac{n\left(t-\mu-\frac{\sigma^2}{n}\right)^2}{2 n \sigma^2}\right) \exp \left(\frac{\frac{\sigma^4}{n}+2 \mu \sigma^2}{2 n \sigma^2}\right) dt\\
&=\exp \left(\frac{\mu}{n}+\frac{\sigma^2}{2 n^2}\right) \int_{-\infty}^{\infty} \frac{1}{\sigma \sqrt{2 \pi}} \exp \left(-\frac{\left(t-\mu-\frac{\sigma^2}{n}\right)^2}{2  \sigma^2}\right) d t , \numberthis\label{eq:gaussian_kernel}
\end{align*}
where (\ref{eq:change_of_variable}) uses the change of variable technique, i.e., $x=\exp(t)$. Note that $\frac{1}{\sigma \sqrt{2 \pi}} \exp \left(-\frac{\left(t-\mu-\frac{\sigma^2}{n}\right)^2}{2  \sigma^2}\right)$ in (\ref{eq:gaussian_kernel})is the pdf of a normal distribution with mean $\mu+\frac{\sigma^2}{n}$ variance $\sigma^2$, thus $\int_{-\infty}^{\infty} \frac{1}{\sigma \sqrt{2 \pi}} \exp \left(-\frac{\left(t-\mu-\frac{\sigma^2}{n}\right)^2}{2  \sigma^2}\right) d t = 1$, and therefore, $\E [X_i^{*\frac{1}{n}}]=\exp \left(\frac{\mu}{n}+\frac{\sigma^2}{2 n^2}\right)$.

The result of (3) follows from
\begin{align*}
\E[\textrm{GM}(X^*)] = \E[\prod_{i=1}^nX^*_i)^{\frac{1}{n}}]= \prod_{i=1}^n\E[X^*_i)^{\frac{1}{n}}]=\prod_{i=1}^n\exp \left(\frac{\mu}{n}+\frac{\sigma^2}{2 n^2}\right)=\exp \left(\mu+\frac{\sigma^2}{2 n}\right).
\end{align*}
\end{proof}
\section{Proof of Theorem \ref{thm:IUT_level}}
\begin{proof}
Let $\theta\in\bigcup_{i=1}^k\Theta_i$. Then $\theta \in \Theta_i$, for some $i=1, \ldots, k$ and
\begin{align*}
\mathbb{P}_\theta(\bm{X}\in R)=\mathbb{P}_\theta(\bm{X}\in \bigcap_{i=1}^kR_i)\leq \mathbb{P}_\theta(\bm{X}\in R_i)\leq\alpha_i\leq\sup_{i=1, \ldots,k}\alpha_i=\alpha.
\end{align*}
Since the above equation holds true for arbitrary $\theta\in\bigcup_{i=1}^k\Theta_i$, the defined IUT test is a level $\alpha$ test.
\end{proof}
\section{Proof of Theorem \ref{thm:IUT_size}}
\begin{proof}
By Theorem \ref{thm:IUT_level}, we know rejection region $R=\bigcap_{j=1}^kR_j$ yields a level-$\alpha$ test, that is
\begin{equation} \label{eq:size_upperbound}
\sup_{\theta\in \bigcup_{j=1}^k\Theta_j}\mathbb{P}_{\theta}(\bm{X}\in R)\leq \alpha.
\end{equation}
Next, we are going to show that $\sup_{\theta\in \bigcup_{j=1}^k\Theta_j}\mathbb{P}_{\theta}(\bm{X}\in R)\geq \alpha$. Because $\theta_i\in\bigcup_{j=1}^k\Theta_j$, therefore,
\begin{align*}
\sup_{\theta\in \bigcup_{j=1}^k\Theta_j}\mathbb{P}_{\theta}(\bm{X}\in R)&\geq \lim_{l\rightarrow \infty}\mathbb{P}_{\theta_l}(\bm{X}\in R)\\
&=\lim_{l\rightarrow \infty}\mathbb{P}_{\theta_l}(\bm{X}\in \bigcap_{j=1}^kR_j)\\
&\geq \lim_{l\rightarrow \infty}\sum_{j=1}^k\mathbb{P}_{\theta_l}(\bm{X}\in R_j) - k + 1,
\end{align*}
where the last inequality follows from Bonferroni's inequality. By condition (1) and (2), we can get $\sum_{j=1}^k\mathbb{P}_{\theta_l}(\bm{X}\in R_j) = k-1+\alpha$. Thus, \begin{equation} \label{eq:size_lowerbound}
\sup_{\theta\in \bigcup_{j=1}^k\Theta_j}\mathbb{P}_{\theta}(\bm{X}\in R)\geq k-1+\alpha-k+1=\alpha.
\end{equation}
Combine (\ref{eq:size_upperbound}) and (\ref{eq:size_lowerbound}), we prove that the size of IUT with rejection region $R=\bigcap_{j=1}^kR_j$ is exactly $\alpha$.
\end{proof}
\section{Proof of Theorem \ref{thm:TOST_size}}
\begin{proof}
Because TOST belongs to IUT, by Theorem \ref{thm:IUT_level}, we know the size of TOST is at most $\alpha$. Consider a parameter point ${\theta_l}=\mu_T - \mu_R = \theta_U$, then we have
\begin{equation*}
\lim_{l\rightarrow \infty}\mathbb{P}_{\theta_l}(\bm{X}\in R_2) = \mathbb{P}_{\theta_U}(\bm{X}\in R_2) = \mathbb{P}_{\theta_U}(T_U < -t_{1-\alpha, r})=\alpha,
\end{equation*}
Furthermore,
\begin{equation*}
\lim_{l\rightarrow \infty}\mathbb{P}_{\theta_l}(\bm{X}\in R_1) = \mathbb{P}_{\theta_U}(\bm{X}\in R_1)= \mathbb{P}_{\theta_U}(T_L > t_{1-\alpha, r}) \rightarrow 1, \textrm{ as }\sigma^2 \rightarrow 0,
\end{equation*}
therefore, the conditions in Theorem \ref{thm:IUT_size} hold true. Thus, the size of TOST is exactly $\alpha$.
\end{proof}
\section{Proof of Theorem \ref{thm:confidence_set_test}}
\begin{proof}
By the definition of acceptance region, we have
\begin{align*}
\sup_{\theta=\theta_0}\mathbb{P}(\bm{X}\notin A(\theta_0)) = \sup_{\theta=\theta_0}\mathbb{P}(T_{\theta_0}=1)\leq\alpha,
\end{align*}
which is the same as
\begin{align*}
1-\alpha\leq\inf_{\theta=\theta_0}\mathbb{P}(\bm{X}\in A(\theta_0)) = \inf_{\theta=\theta_0}\mathbb{P}(\theta_0\in C(\bm{X})).
\end{align*}
Since the above statements hold true for all $\theta\in\Theta$, thus
\begin{align*}
\inf_{\theta_0\in\Theta}\inf_{\theta=\theta_0}\mathbb{P}(\theta_0\in C(\bm{X})) \geq 1-\alpha,
\end{align*}
which implies $C(\bm{X})=\{\theta: \bm{X}\in A(\theta)\}$ is a level-$\alpha$ confidence set for $\theta$.
\end{proof}
\section{Proof of Theorem \ref{thm:confidence_interval_alpha}}
\begin{proof}
If $\mu_T - \mu_R = 0$, then $\mu_T - \mu_R \in [L(\bm{X}), U(\bm{X})]$ for all $\bm{X}$, thus the coverage probability of $[L(\bm{X}), U(\bm{X})]$ is 1 in this case. So it is only need to consider the case when $\mu_T - \mu_R \neq 0$.

When $\mu_T - \mu_R > 0$, the event $\{\mu_T - \mu_R \geq \min\{0, (\bar{X}_T - \bar{X}_R) - t_{1-\alpha, r}\hat{\sigma}_{\Delta\bar{X}}\}\}$ is the whole sample space, thus
\begin{align*}
\mathbb{P}\left(\{\mu_T - \mu_R \geq \min\{0, (\bar{X}_T - \bar{X}_R) - t_{1-\alpha, r}\hat{\sigma}_{\Delta\bar{X}}\}\}\right)=1.
\end{align*}
And if $\mu_T - \mu_R > 0$, the event $\{\mu_T - \mu_R \leq \max\{0, (\bar{X}_T - \bar{X}_R) + t_{1-\alpha, r}\hat{\sigma}_{\Delta\bar{X}}\}\}$ is the same as event
$\left\{\mu_T - \mu_R \leq (\bar{X}_T - \bar{X}_R) + t_{1-\alpha, r}\hat{\sigma}_{\Delta\bar{X}}\right\}$, so the coverage probability equals
\begin{align*}
\mathbb{P}\left(\mu_T - \mu_R \in [L(\bm{X}), U(\bm{X})]\right) &= \mathbb{P}\left(\mu_T - \mu_R \leq \max\{0, (\bar{X}_T - \bar{X}_R) + t_{1-\alpha, r}\hat{\sigma}_{\Delta\bar{X}}\}\right)\\
& = \mathbb{P}\left(\mu_T - \mu_R \leq (\bar{X}_T - \bar{X}_R) + t_{1-\alpha, r}\hat{\sigma}_{\Delta\bar{X}}\right)\\
& = \mathbb{P}\left(\bar{X}_T - \bar{X}_R- (\mu_T - \mu_R) \geq -t_{1-\alpha, r}\hat{\sigma}_{\Delta\bar{X}}\right)\\
&=1-\alpha.
\end{align*}

Similarly, when $\mu_T - \mu_R < 0$, the event $\{\mu_T - \mu_R \leq \max\{0, (\bar{X}_T - \bar{X}_R) + t_{1-\alpha, r}\hat{\sigma}_{\Delta\bar{X}}\}\}$ is the whole sample space, thus
\begin{align*}
\mathbb{P}\left(\{\mu_T - \mu_R \leq \max\{0, (\bar{X}_T - \bar{X}_R) + t_{1-\alpha, r}\hat{\sigma}_{\Delta\bar{X}}\}\}\right)=1.
\end{align*}
If $\mu_T - \mu_R < 0$, the event $\{\mu_T - \mu_R \geq \min\{0, (\bar{X}_T - \bar{X}_R) - t_{1-\alpha, r}\hat{\sigma}_{\Delta\bar{X}}\}\}$ is the same as event
$\left\{\mu_T - \mu_R \geq (\bar{X}_T - \bar{X}_R) - t_{1-\alpha, r}\hat{\sigma}_{\Delta\bar{X}}\right\}$, so the coverage probability equals
\begin{align*}
\mathbb{P}\left(\mu_T - \mu_R \in [L(\bm{X}), U(\bm{X})]\right) &= \mathbb{P}\left(\mu_T - \mu_R \geq \min\{0, (\bar{X}_T - \bar{X}_R) - t_{1-\alpha, r}\hat{\sigma}_{\Delta\bar{X}}\}\right)\\
& = \mathbb{P}\left(\mu_T - \mu_R \geq (\bar{X}_T - \bar{X}_R) - t_{1-\alpha, r}\hat{\sigma}_{\Delta\bar{X}}\right)\\
& = \mathbb{P}\left(\bar{X}_T - \bar{X}_R- (\mu_T - \mu_R) \leq t_{1-\alpha, r}\hat{\sigma}_{\Delta\bar{X}}\right)\\
&=1-\alpha.
\end{align*}
Combine everything above, the proof is completed.
\end{proof}
\section{Proof of Theorem \ref{thm:confidence_interval_2alpha}}
\begin{proof}
By definition, we have
\begin{equation*}
\mathbb{P}(L(\bm{X})\leq \theta) = 1- \alpha_1, \mathbb{P}(U(\bm{X}) \geq \theta) = 1-\alpha_2.
\end{equation*}
Consider the events $\mathcal{A}$ and $\mathcal{B}$ defined as $\mathcal{A} = \{\bm{X}: L(\bm{X})\leq \theta\}, \mathcal{B} = \{\bm{X}: U(\bm{X})\geq \theta\}$. Therefore, the interaction of $\mathcal{A}$ and $\mathcal{B}$ is $\mathcal{A} \bigcap \mathcal{B} = \{\bm{X}: L(\bm{X})\leq \theta \leq U(\bm{X})\}$. It follows that
\begin{equation*}
\mathbb{P}\left(\mathcal{A} \bigcup \mathcal{B}\right) =  \mathbb{P}\left(L(\bm{X})\leq \theta \textrm{ or } U(\bm{X})\geq\theta\right)\geq
\mathbb{P}\left(L(\bm{X})\leq \theta \textrm{ or } L(\bm{X})\geq\theta\right) =1.
\end{equation*}
Since $\mathbb{P}\left(\mathcal{A} \bigcup \mathcal{B}\right)\leq 1$, we have $\mathbb{P}\left(\mathcal{A} \bigcup \mathcal{B}\right) = 1$. The result follow from that
\begin{equation*}
\mathbb{P}\left(\mathcal{A} \bigcap \mathcal{B}\right) = \mathbb{P}(\mathcal{A}) + \mathbb{P}(\mathcal{B}) - \mathbb{P}\left(\mathcal{A} \bigcup \mathcal{B}\right) = 1- \alpha_1 - \alpha_2.
\end{equation*}
\end{proof}

\end{document}